\documentclass[11pt]{article}
\pdfoutput=1
\usepackage{color}
\usepackage[textwidth = 430 pt, textheight = 630 pt]{geometry}
\definecolor{MyDarkBlue}{rgb}{0.15,0.25,0.45}
\usepackage{epsfig,rotating}
\usepackage{amsmath,amssymb}
\usepackage{amsfonts}
\usepackage{mathrsfs}
\usepackage{bbm}
\usepackage{bm}
\usepackage{hyperref}
\hypersetup{
colorlinks=true,
citecolor=MyDarkBlue,
linkcolor=MyDarkBlue,
urlcolor=MyDarkBlue,
pdfauthor={Sam Palmer and Christian S\"amann},
pdftitle={M-brane Models from Non-Abelian Gerbes},
pdfsubject={hep-th}
breaklinks=true
}

\let\fn\footnote
\renewcommand{\footnote}[1]{\linespread{1.1}\fn{#1}\linespread{1.29}}

\makeatletter\renewcommand{\section}{\@startsection
{section}{1}{\z@}{-3.5ex plus -1ex minus
    -.2ex}{2.3ex plus .2ex}{\bf }}
\makeatletter\renewcommand{\subsection}{\@startsection{subsection}{2}{\z@}{-3.25ex
plus -1ex minus
   -.2ex}{1.5ex plus .2ex}{\it }}
\makeatletter\renewcommand{\subsubsection}{\@startsection{subsubsection}{3}{-2.45ex}{-3.25ex
plus -1ex minus -.2ex}{1.5ex plus .2ex}{\it }}
\renewcommand{\thesection}{\arabic{section}}
\renewcommand{\thesubsection}{\arabic{section}.\arabic{subsection}}
\renewcommand{\@seccntformat}[1]{\@nameuse{the#1}.~~}

\renewcommand{\theequation}{\thesection.\arabic{equation}}
\makeatletter \@addtoreset{equation}{section}

\renewenvironment{thebibliography}[1]
     {\baselineskip=16pt plus 2pt minus 1pt
      \section*{\large\refname
        \@mkboth{\MakeUppercase\refname}{\MakeUppercase\refname}}%
     \list{\@biblabel{\@arabic\c@enumiv}}%
           {\settowidth\labelwidth{\@biblabel{#1}}%
            \leftmargin\labelwidth
            \advance\leftmargin\labelsep
            \@openbib@code
            \usecounter{enumiv}%
            \let\p@enumiv\@empty
            \renewcommand\theenumiv{\@arabic\c@enumiv}}%
      \sloppy
      \clubpenalty4000
      \@clubpenalty \clubpenalty
      \widowpenalty4000%
      \sfcode`\.\@m}

\newcommand{\appendices}{
\section*{Appendix}\label{appendices}\setcounter{subsection}{0}
\addcontentsline{toc}{section}{Appendix}
\setcounter{equation}{0}
\makeatletter
\renewcommand{\theequation}{\Alph{subsection}.\arabic{equation}}
\renewcommand{\thesubsection}{\Alph{subsection}}
\@addtoreset{equation}{subsection}
\makeatother
}

\def\periodb#1{\setbox0=\hbox{$#1$}#1\hskip-\wd0\hbox to\wd0{-}}

\newcommand{\lbr}{(\hspace{-0.1cm}(}
\newcommand{\rbr}{)\hspace{-0.1cm})}

\newcommand{\unit}{\mathbbm{1}}   			

\newcommand{\CA}{\mathcal{A}}    			

\newcommand{\CC}{\mathcal{C}}

\newcommand{\CF}{\mathcal{F}}

\newcommand{\CG}{\mathcal{G}}

\newcommand{\CH}{\mathcal{H}}

\newcommand{\CL}{\mathcal{L}}

\newcommand{\CN}{\mathcal{N}}

\newcommand{\CE}{\mathcal{E}}
\newcommand{\frg}{\mathfrak{g}}				
\newcommand{\frh}{\mathfrak{h}}				

\newcommand{\FR}{\mathbbm{R}}     			
\newcommand{\FC}{\mathbbm{C}}     			

\newcommand{\dd}{\mathrm{d}}     			
\newcommand{\dpar}{\partial}     			
\newcommand{\di}{\mathrm{i}}     			
\newcommand{\eps}{{\varepsilon}}			
\newcommand{\epsb}{{\bar{\varepsilon}}}			

\newcommand{\eand}{{~~~\mbox{and}~~~}}     		

\newcommand{\der}[1]{\frac{\dpar}{\dpar #1}}   		
\newcommand{\dder}[1]{\frac{\dd}{\dd #1}}   		
\newcommand{\tr}{\,\mathrm{tr}\,}     			

\newcommand{\agl}{\mathfrak{gl}}     			

\newcommand{\au}{\mathfrak{u}}
\newcommand{\asu}{\mathfrak{su}}
\newcommand{\aso}{\mathfrak{so}}

\newcommand{\sU}{\mathsf{U}}     			

\newcommand{\actwedge}{{\scriptstyle\,\,\stackrel{\acton}{\wedge}\,\,}}

\newcommand{\acton}{\vartriangleright}     			
\newcommand{\remark}[1]{}     				
\def\tyng(#1){\hbox{\tiny$\yng(#1)$}}			
\def\tyoung(#1){\hbox{\tiny$\young(#1)$}}			

\newcommand{\mt}{\mathsf{t}}
\newcommand{\sft}{\mathsf{t}}
\newcommand{\eor}{{~~~\mbox{or}~~~}}

\begin{document}

\begin{titlepage}
\begin{flushright}
 HWM--12--03 \\ EMPG--12--05
\end{flushright}
\vskip 2.0cm
\begin{center}
{\LARGE \bf M-brane Models from Non-Abelian Gerbes}
\vskip 1.5cm
{\Large Sam Palmer and Christian S\"amann}
\setcounter{footnote}{0}
\renewcommand{\thefootnote}{\arabic{thefootnote}}
\vskip 1cm
{\em Department of Mathematics and \\
Maxwell Institute for Mathematical Sciences\\[0.1cm]
Heriot-Watt University\\
Colin Maclaurin Building, Riccarton, Edinburgh EH14 4AS, U.K.}\\[0.5cm]
{Email: {\ttfamily sap2@hw.ac.uk~,~c.saemann@hw.ac.uk}}
\end{center}
\vskip 1.0cm
\begin{center}
{\bf Abstract}
\end{center}
\begin{quote}
We make the observation that M-brane models defined in terms of 3-algebras can be interpreted as higher gauge theories involving Lie 2-groups. Such gauge theories arise in particular in the description of non-abelian gerbes. This observation allows us to put M2- and M5-brane models on equal footing, at least as far as the gauge structure is concerned. Furthermore, it provides a useful framework for various generalizations; in particular, it leads to a fully supersymmetric generalization of a previously proposed set of tensor multiplet equations.
\end{quote}
\end{titlepage}

\section{Introduction and results}

A recently proposed generalized Nahm transform \cite{Saemann:2010cp,Palmer:2011vx}, see also \cite{Gustavsson:2008dy}, connects solutions of the Basu-Harvey equation \cite{Basu:2004ed} to solutions of a loop space version of the self-dual string equation \cite{Howe:1997ue}. This is somewhat surprising, as the original Nahm transform is a duality between identical equations: instanton solutions on a four-torus are mapped to instantons on the corresponding dual four-torus. Taking certain infinite-radius limits, one then arrives e.g.\ at the ADHMN construction of solutions to the Bogomolny monopole equation from solutions to the Nahm equation. The Basu-Harvey equation and the self-dual string equation, however, seem very different. If a full Nahm transform is to exist, one would need a reformulation of both equations in a common language.

A non-abelian version of the self-dual string equation is known only on loop space \cite{Saemann:2010cp,Palmer:2011vx,Gustavsson:2008dy}. However, it seems clear that a direct formulation on space-time is very likely to involve the non-abelian gerbes defined in \cite{Breen:math0106083} and equivalently in \cite{Aschieri:2003mw}. These non-abelian gerbes can be described in terms of higher gauge theories involving Lie 2-groups as gauge groups. The category of (strict) Lie 2-groups is equivalent to the category of Lie crossed modules and the gauge algebra in the higher gauge theories is therefore given by differential crossed modules. In this letter, we make the observation that 3-algebras relevant to M-brane models as discussed in \cite{Bagger:2008se,Cherkis:2008qr}, see also \cite{deMedeiros:2008zh}, are special cases of differential crossed modules. Therefore, these models can be regarded as higher gauge theories.

The existence of our reformulation can also be expected from the following point of view: the fuzzy funnel described by the Basu-Harvey equation \cite{Basu:2004ed} should contain a non-commutative 3-sphere. One would expect that the corresponding Hilbert space is a categorification of an ordinary Hilbert space and therefore based on a 2-vector space. The fields in the Basu-Harvey equation should correspond to endomorphisms of this 2-Hilbert space, which are organized in the structure of a Lie 2-algebra.

Besides making the existence of a full generalized Nahm transform more conceivable, our reformulation also has other advantages. In particular, it yields more than the 3-algebra based M-brane models already known, and therefore we can use it as a framework\footnote{Another framework for generalizing the Basu-Harvey equation and M2-brane models in general has been proposed in \cite{IuliuLazaroiu:2009wz} in the form of strong homotopy Lie algebras, or $L_\infty$-algebras for short. The $L_\infty$-structures identified there are different from the ones found here.} for generalizing these models. Within this larger class of models, one might overcome some of the problems of current M-brane models. For example, one might hope to find Chern-Simons matter theories with $\CN=8$ supersymmetry beyond the Bagger-Lambert-Gustavsson model based on the 3-Lie algebra $A_4$. Encouraging is that we do find new $\CN=(2,0)$ supersymmetric tensor multiplet equations beyond a recently proposed set based on 3-Lie algebras.

A detailed study of the implications of our observation is beyond the scope of this letter, and we will focus our discussion of M-brane models mainly on the BLG model \cite{Bagger:2007jr,Gustavsson:2007vu}, the ABJM model \cite{Aharony:2008ug,Bagger:2008se} and a set of equations proposed for a 3-Lie algebra valued tensor multiplet \cite{Lambert:2010wm}. We will only employ strict Lie 2-groups and work with local non-abelian gerbes. The matter fields will be restricted to sections of an associated principal\linebreak (1-)bundle of the gauge group. In future work, we plan to extend our discussion beyond these limitations. Furthermore, we intend to study supersymmetric Chern-Simons matter theories as well as the quantization of $S^3$ in our reformulation. A detailed comparison with other M-brane models, e.g.\  the ones proposed in \cite{Ho:2011ni,Samtleben:2011fj,Chu:2011fd,Fiorenza:2012tb}, would also be interesting.

We start our discussion by reviewing Lie 2-groups and crossed modules as well as the derivation of 3-algebras from these. We give a few non-trivial examples that go beyond the usual picture of 3-algebras. We then present the interpretation of the 3-Lie algebra valued tensor multiplet equations of \cite{Lambert:2010wm} as well as the M2-brane models of \cite{Bagger:2007jr,Gustavsson:2007vu,Aharony:2008ug,Bagger:2008se} in the framework of higher gauge theories.

\section{Lie 2-algebras and 3-algebras}

In this section, we make the connection between Lie 2-algebras and 3-algebras using an extension of the so-called Faulkner construction. For a detailed account of the Faulkner construction for 3-algebras, see \cite{deMedeiros:2008zh}. For a motivation and a more extensive discussion of categorified gauge structures, we refer to \cite{Baez:2002jn,Baez:0511710,Baez:2010ya}.

\subsection{Lie 2-groups, Lie 2-algebras and crossed modules}

While the parallel transport of a point particle along a path assigns a group element to each path, the parallel transport of a string along a surface leads naturally to the concept of a Lie 2-group\footnote{In this letter, we will restrict ourselves to strict Lie 2-groups.}. A Lie 2-group is a categorification of the notion of a Lie group. Recall that a group is a (small) category with one object in which each morphism is invertible. A Lie 2-group is analogously built from a corresponding 2-category, i.e.\ a category with additional ``morphisms between morphisms''. Furthermore, the category of Lie 2-groups can be shown to be equivalent to the category of Lie crossed modules and it is this language that we will use.

Recall that a {\em crossed module} is a pair of groups $\CG$ and $\CH$ together with an automorphism action $\acton$ of $\CG$ onto $\CH$ and a group homomorphisms $\mt:\CH\rightarrow \CG$, which satisfy the following conditions:
\begin{itemize}
\item[$i)$] $\mt$ is equivariant with respect to conjugation,
\begin{subequations}\label{eq:LCMIdentities}
\begin{equation}
 \mt(g\acton h)=g \mt(h) g^{-1}~,
\end{equation}
\item[$ii)$] and the so-called {\em Pfeiffer identity} holds:
\begin{equation}
 \mt(h_1)\acton h_2=h_1h_2h_1^{-1}~,
\end{equation}
\end{subequations}
\end{itemize}
for all $g\in\CG$ and $h,h_1,h_2\in\CH$. A {\em Lie crossed module} is a crossed module $(\mt:\CH\rightarrow\CG,\acton)$, where $\CG$ and $\CH$ are Lie groups. A simple example of a Lie crossed module is $\CG=\CH=\sU(N)$ with $\mt$ the identity map and $\acton$ the adjoint action.

Just as a Lie algebra can be obtained by linearizing a Lie group at the identity element, so can a Lie 2-algebra be obtained by linearizing a Lie 2-group. These Lie 2-algebras correspond to differential crossed modules.

A {\em differential crossed module} $(\sft:\frh\rightarrow \frg,\acton)$ is a pair of Lie algebras $\frg$, $\frh$ together with an action $\acton$ of elements of $\frg$ as derivations of $\frh$ and a Lie algebra homomorphism between $\frh$ and $\frg$, which we will also denote by $\mt$, slightly abusing notation. We demand that $\acton$ and $\mt$ satisfy the linearized versions of the identities \eqref{eq:LCMIdentities}:
\begin{equation}
 \mt(x\acton y)=[x,\mt(y)]\eand \mt(y_1)\acton y_2=[y_1,y_2]
\end{equation}
for all $x\in \frg$ and $y,y_1,y_2\in \frh$. The differential version of our simple example from above is evidently $\frg=\frh=\au(N)$ with $\acton$ being the adjoint action and $\sft$ the identity map.

Note that Lie 2-algebras are 2-term $L_\infty$-algebras \cite{Baez:2003aa}. These have a 3-bracket called the Jacobiator, which is different from the 3-bracket we define later. These 2-term $L_\infty$-algebras are only equivalent to differential crossed modules when the Jacobiator vanishes. In this case, they are called \emph{strict} Lie 2-algebras.

To write down action functionals, we need to extend the above notion to that of a {\em metric differential crossed module}. The additional metric structure on a differential crossed module $(\mt:\frh\rightarrow\frg,\acton)$ is given by non-degenerate hermitian forms $\lbr\cdot,\cdot\rbr$ on $\frg$ and $(\cdot,\cdot)$ on $\frh$, which are invariant under the obvious Lie algebra actions:
\begin{equation}
\begin{aligned}
 \lbr [x_1,x_2],x_3\rbr+\lbr x_2,[\bar x_1,x_3]\rbr&=0~,\\
 (x\acton y_1,y_2)+(y_1,\bar x\acton y_2)&=0~.
\end{aligned}
\end{equation}
The last equation also implies that $(\cdot,\cdot)$ is $\frh$-invariant: $ ([y_1,y_2],y_3)+(y_2,[\bar y_1,y_3])=0$.

Note that the introduction of the metric structure allows us to define a map $\mt^*:\mathfrak g\rightarrow\mathfrak h$ implicitly by
\begin{equation}
 (\mt^*(x),y):=\lbr x,\mt(y)\rbr~.
\end{equation}
One readily verifies useful identities, e.g.\
\begin{equation}\label{eq:tstaridentity}
 \sft^*([x_1,x_2])=x_1\acton \sft^*(x_2)=-x_2\acton \sft^*(x_1)~.
\end{equation}

To avoid the appearance of ghosts from matter fields in our M-brane models, we will always choose the metric on $\frh$ to be positive definite. For $\frg$, however, we would like to allow split signature. The reason for this is that all the 3-algebra M2-brane models are given by Chern-Simons matter theories, which are a priori not parity invariant. Having a gauge algebra $\frg$ of the form $\frg_L\oplus \frg_R$ with split signature yields a pair of Chern-Simons terms with opposite Chern-Simons levels. These are then mapped into each other under a parity flip.

\subsection{3-algebras}

In the M-theory generalization of the Nahm equation proposed by Basu and Harvey \cite{Basu:2004ed}, Filippov's 3-Lie algebras \cite{Filippov:1985aa} play a prominent role. A {\em 3-Lie algebra} is a real vector space $\CA$ endowed with a totally antisymmetric, trilinear map $[\cdot,\cdot,\cdot]:\CA^{\wedge 3}\rightarrow \CA$, which satisfies the so-called {\em fundamental identity}:
\begin{equation}\label{eq:fun}
 [a_1,a_2,[b_1,b_2,b_3]]=[[a_1,a_2,b_1],b_2,b_3]+[b_1,[a_1,a_2,b_2],b_3]+[b_1,b_2,[a_1,a_2,b_3]]
\end{equation}
for all $a_1,a_2,b_1,b_2,b_3\in \CA$. Due to this identity, the span of the operators $D(a,b)$, $a,b\in\CA$, which act on $c\in\CA$ according to
\begin{equation}
 D(a,b)\acton c:=[a,b,c]~,
\end{equation}
forms a Lie algebra. We will call this Lie algebra the {\em associated Lie algebra} of $\CA$ and denote it $\frg_\CA$. We can turn $\CA$ into a {\em metric 3-Lie algebra} by introducing a positive definite, non-degenerate symmetric bilinear form $(\cdot,\cdot)$ on $\CA$, which is invariant under actions of $\frg_\CA$:
\begin{equation}
 ([a_1,a_2,b_1],b_2)+(b_1,[a_1,a_2,b_2])=0~.
\end{equation}

Because the only 3-Lie algebras with positive definite metric are $A_4$ (which is the four-dimensional 3-Lie algebra ${\rm span}(e^\mu)$, $\mu=1,\ldots,4$, with 3-bracket $[e^\mu,e^\nu,e^\kappa]=\eps^{\mu\nu\kappa\lambda}e^\lambda$) and direct sums thereof \cite{Nagy:2007aa}, generalizations of the above 3-Lie algebras were soon proposed. Here, we will drop the total antisymmetry (and the word Lie from 3-Lie algebra) and focus on the real and hermitian 3-algebras introduced in \cite{Cherkis:2008qr} and \cite{Bagger:2008se}. Real (and hermitian) 3-algebras are (complex) vector spaces endowed with 3-brackets which are anti-symmetric in their first two slots. The 3-brackets are linear in all their slots except for the third slot of hermitian 3-algebras, which is antilinear. They are required to satisfy the fundamental identity and the metric compatibility condition, which can be written in an intuitive form if we define
\begin{equation}\label{eq:3bra}
[a_1,a_2,a_3]:=D(a_1,a_2)\acton a_3~~\mbox{and}~~[a_3,a_1;a_2]:=D(a_1,a_2)\acton a_3~,
\end{equation}
for real and hermitian 3-algebras, respectively. We also define a complex conjugation $\overline{D(a_1,a_2)}:=-D(a_2,a_1)$, which leaves the inner derivations of real 3-algebras invariant. For both real and hermitian 3-algebras, we can then write the fundamental identity as
\begin{equation}\label{eq:GenFundIdent}
[D(a_1,a_2),D(b_1,b_2)]\acton c=D(D(a_1,a_2)\acton b_1,b_2)\acton c+D(b_1,\overline{D(a_1,a_2)}\acton b_2)\acton c
\end{equation}
and the metric compatibility condition as
\begin{equation}
 (D(a_1,a_2)\acton b_1,b_2)+(b_1,\overline{D(a_1,a_2)}\acton b_2)=0~.
\end{equation}

Using the 3-bracket and the metric on the 3-algebra, a nondegenerate invariant metric on the associated Lie algebra is induced by defining
\begin{equation}
 \lbr D(a_1,a_2),D(a_3,a_4)\rbr:=-(D(a_1,a_2)\acton a_4,a_3)~.
\end{equation}
Note that here, we started from a 3-bracket on a metric 3-algebra $\CA$ and constructed the map $D$ and a metric on $\frg_\CA$. In the following subsection, we will perform the inverse operation. We will start from a differential crossed module and construct a map $D$ and hence a 3-bracket.

\subsection{Deriving 3-algebras from differential crossed modules}

It is possible to construct all 3-algebras from metric Lie algebras together with certain faithful representations via the Faulkner construction \cite{Faulkner:1973aa,deMedeiros:2008zh}. These pairs of Lie algebras and representations correspond to metric differential crossed modules $(\mt:\frh\rightarrow\frg,\acton)$ with abelian $\frh$ and trivial $\mt$. Thus, all real and hermitian 3-algebras are obtained by applying the Faulkner construction to such differential crossed modules whose Lie algebras $\frh$ are real or complex, respectively. However, we can extend this construction to arbitrary metric differential crossed modules: Allowing $\frh$ to be non-abelian and $\mt$ non-trivial still gives structures with 3-brackets which satisfy the fundamental identity \eqref{eq:fun}.

Starting from a metric differential crossed module $(\mt:\frh\rightarrow\frg,\acton)$, there is a unique linear map $D: \frh\otimes \frh \rightarrow \frg$ such that\footnote{The usual definition, in e.g.\ \cite{Baez:2002jn}, is $\lbr x,D(y_1,y_2)\rbr = (x\acton y_1,y_2)$. This agrees with our definition in the real case, but in the complex case our definition gives antilinearity in the second argument, which is the convention chosen for the hermitian 3-algebras in \eqref{eq:3bra}.}
\begin{equation}
 \lbr x,D(y_1,y_2)\rbr=-(x\acton y_2,y_1)
\end{equation}
for all $x\in \frg$ and $y_1,y_2\in \frh$. The map $D$ is skew-hermitian since
\begin{equation}
\begin{aligned}
 \lbr x,D(y_1,y_2)\rbr&=-(x\acton y_2,y_1)=(y_2,\bar x\acton y_1)\\&=\overline{(\bar x\acton y_1,y_2)}=-\overline{\lbr \bar x,D(y_2,y_1)\rbr}=\lbr x,-\overline{D(y_2,y_1)}\rbr~,
\end{aligned}
\end{equation}
and satisfies the identity \cite{Martins:2010ry}
\begin{equation}
[x,D(y_1,y_2)]=D(x\acton y_1,y_2)+D( y_1,\bar x\acton y_2)~,
\end{equation}
which implies the fundamental identity \eqref{eq:GenFundIdent}. Therefore, we can define 3-brackets according to
\begin{equation}
[y_1,y_2,y_3] :=D(y_1,y_2)\acton y_3\eand [y_3,y_1;y_2] :=D(y_1,y_2)\acton y_3
\end{equation}
for real and hermitian 3-algebras, respectively.

\subsection{3-algebra examples}

Let us now reconstruct the familiar examples of 3-algebras. The simplest way of realizing the 3-algebra $A_4$ as a differential crossed module is to take $(\mt:\FR^4\rightarrow\aso(4),\acton)$, where $\acton$ is the ordinary action of $\aso(4)$ on the fundamental representation, $\sft=0$ is the trivial map and the metric on $\aso(4)\cong\asu(2)\times\asu(2)$ is of split signature. $\FR^4$ is viewed here as an abelian Lie algebra with trivial Lie bracket and Euclidean metric. This gives the completely anti-symmetric 3-bracket $[e_\mu,e_\nu,e_\kappa]:=D(e_\mu,e_\nu)\acton e_\kappa=\eps_{\mu\nu\kappa\lambda}e_\lambda$ on the standard basis vectors $e_\mu\in\FR^4$.

The hermitian 3-algebras occurring in the ABJM model \cite{Aharony:2008ug,Bagger:2008se} are equivalent to crossed modules of the form $(\mt:\agl(n,\FC)\rightarrow \agl(n,\FC)\times \agl(n,\FC),\acton)$, where $\mt=0$ and $\frh=\agl(n,\FC)$ is regarded as an (additive) abelian Lie algebra. The action of $\frg$ on $\frh$ is given by
\begin{equation}
(x_1,x_2)\acton y := x_1 y  - y  x_2~,
\end{equation}
which yields the following Lie bracket:
\begin{equation}
[(x_1,x_2), (x_3,x_4)]=([x_1,x_3],[x_2,x_4])~.
\end{equation}
The metric structures on $\frh$ and $\frg$ are given by
\begin{equation}
(y_1,y_2):=\tr(y_1 y_2^\dagger)~,~~\lbr (x_1,x_2), (x_3,x_4)\rbr:=\tr(x_1 x_3^\dagger-x_2 x_4^\dagger)~,
\end{equation}
and from these we derive the derivations
\begin{equation}
D(y_1,y_2)= (y_1  y_2^\dagger, y_2^\dagger y_1)~,
\end{equation}
which yield the 3-bracket
\begin{equation}
[y_1,y_3;y_2]:=D(y_1,y_2)\acton y_3= y_1  y_2^\dagger y_3-y_3 y_2^\dagger y_1~.
\end{equation}

The 3-algebras $\CC_{2n}$ used in \cite{Cherkis:2008qr} are\footnote{This can be easily restricted to the real case: $(\mt:\agl(n,\FR)\rightarrow\aso(n)\times\aso(n),\acton)$.}  $(\mt:\agl(n,\FC)\rightarrow\asu(n)\times\asu(n),\acton)$, where $\mt=0$, $\frh=\agl(n,\FC)$ is abelian and the action of $\frg$ on $\frh$ reads as
\begin{equation}
(x_1,x_2)\acton y =  x_1 y  - y  x_2~.
\end{equation}
The metrics are
\begin{equation}
(y_1,y_2):=\tr(y_1 y_2^\dagger+ y_1^\dagger y_2)\eand\lbr (x_1,x_2), (x_3,x_4)\rbr:=-\tr(x_1 x_3-x_2 x_4)~,
\end{equation}
from which we derive
\begin{equation}
D(y_1,y_2)= (y_1  y_2^\dagger  - y_2  y_1^\dagger, y_2^\dagger y_1 -  y_1^\dagger y_2)~.
\end{equation}
In the case $n=2$, the bracket is totally anti-symmetric and the 3-algebra becomes $A_4$.

Similarly, one can obtain all 3-algebras, in particular those appearing in the classification of \cite{Cherkis:2008ha}, from differential crossed modules with $\sft=0$.

\subsection{Nontrivial examples of differential crossed modules}\label{sec:NontrivialExamples}

The non-abelian gerbes of Breen and Messing \cite{Breen:math0106083} use automorphism Lie 2-groups, whose differential crossed modules are of the form $(\mt:\frh\rightarrow \mathsf{Der}(\frh),\acton)$, where $\sft$ is the obvious map from the Lie algebra $\frh$ to its derivations $\mathsf{Der}(\frh)$ and $\acton$ is the action of these derivations. The simplest example is $(\mt:\au(n)\rightarrow\au(n),\acton)$, with $\mt$ being the identity and $\acton$ the adjoint action. With Hilbert-Schmidt metrics, this non-abelian gerbe has a 3-bracket
\begin{equation}
[y_1,y_2,y_3]:=D(y_1,y_2)\acton y_3= [[y_1 , y_2], y_3]~.
\end{equation}
This example trivially reduces to a differential crossed module $(\mt:\au(n)\rightarrow\asu(n),\acton)$, where $\sft(\unit):=0$. It is this differential crossed module that we will encounter in the M5-brane model.

Finally, we will consider an example from \cite{Martins:2010ry}. Let $\frh$ be the Lie algebra of complex block matrices with blocks of sizes
\begin{equation}
\begin{pmatrix} m\times m & m\times p\\n\times m & n\times p\end{pmatrix}
\end{equation}
endowed with the Lie bracket (which is not the ordinary matrix commutator)
\begin{equation}
\left[   \begin{pmatrix} A & B \\C  & D\end{pmatrix} ,\begin{pmatrix} A' & B' \\C'  & D'\end{pmatrix} \right]=\begin{pmatrix} [A,A'] &  AB'-A'B \\ CA'-C'A & CB'-C'B\end{pmatrix} ~.
\end{equation}
The Lie algebra $\frg$ consists of pairs of these matrices of the form
\begin{equation}
\left( \begin{pmatrix} A & 0\\C & D\end{pmatrix},\begin{pmatrix} A & B'\\0 & D'\end{pmatrix}  \right)~,
\end{equation}
where the Lie bracket is the usual matrix commutator. Now the map  $\mt:\frh\rightarrow\frg$ is given by
\begin{equation}
\mt\begin{pmatrix} A & B \\C  & D\end{pmatrix}=\left ( \begin{pmatrix} A & 0\\ C & 0 \end{pmatrix}, \begin{pmatrix} A & B\\  0 & 0 \end{pmatrix}\right)~,
\end{equation}
and the action of $\frg$ on $\frh$ is
\begin{equation}
 \left  ( \begin{pmatrix} A & 0\\C & D\end{pmatrix},\begin{pmatrix} A & B'\\0 & D'\end{pmatrix}  \right) \acton \begin{pmatrix} A_1& B_1\\C_1& D_1\end{pmatrix}= \begin{pmatrix} A & 0\\C & D\end{pmatrix} \begin{pmatrix}  A_1& B_1\\C_1& D_1\end{pmatrix} -\begin{pmatrix}  A_1& B_1\\C_1& D_1\end{pmatrix}\begin{pmatrix} A & B'\\0 & D'\end{pmatrix}~.
\end{equation}
We can endow the Lie algebras $\frh$ and $\frg$ with Hilbert-Schmidt metrics, which we choose to be positive definite on $\frh$ and of split signature on $\frg$. Then we find
\begin{equation}
\begin{aligned}
&D\left( \begin{pmatrix} A_1 & B_1 \\C_1  & D_1\end{pmatrix} ,\begin{pmatrix} A_2 & B_2 \\C_2  & D_2\end{pmatrix}\right)\\
&=\left( \begin{pmatrix} A_1 A_2^\dagger+(B_1 B_2^\dagger+C_1 C_2^\dagger)/2 & 0 \\C_1 A_2^\dagger +B_1A_2^\dagger  & C_1C_2^\dagger+D_1D_2^\dagger\end{pmatrix} ,\right.\\&\hspace{4cm}\left.\begin{pmatrix} A_1 A_2^\dagger+(B_1 B_2^\dagger+C_1 C_2^\dagger)/2& C_2^\dagger D_1 +A_2^\dagger B_1  \\0 & B_2^\dagger B_1+D_2^\dagger D_1\end{pmatrix}\right)~,
\end{aligned}
\end{equation}
from which one can derive a corresponding 3-bracket as $[x,y,z]:=D(x,y)\acton z$, where $x,y,z\in\frh$.

\section{M-brane models}

Let us now apply our observation that 3-Lie algebras are special cases of differential crossed modules. After briefly reviewing higher gauge theories, we rewrite a recently proposed set of supersymmetric equations of motion for the non-abelian (2,0) tensor multiplet in this language. We then consider the corresponding re-interpretation of the BLG model.

\subsection{Higher gauge theory with differential crossed modules}

In this letter, we will restrict ourselves to trivial principal 2-bundles over $\FR^n$, such that there is no distinction between local and global objects. Similar to trivial principal bundles, all \v{C}ech cocycles defining the bundle are trivial, and all non-trivial information is contained in the connection. Moreover, all potentials defining this connection are given in terms of Lie algebra valued differential forms.

Consider a (trivial) principal 2-bundle $\CE$ over $\FR^n$. Let the structure Lie 2-group of $\CE$ be given in terms of the Lie crossed module $(\sft:\CH\rightarrow \CG,\acton)$ with corresponding differential crossed module $(\sft:\frh\rightarrow \frg,\acton)$. A {\em connection} on $\CE$ is a pair $(A,B)$, where $A$ is a $\frg$-valued 1-form and $B$ is an $\frh$-valued 2-form, cf.\ e.g.\ \cite{Baez:2002jn}. We also introduce the corresponding {\em curvatures} as a pair $(F,H)$, where $F$ takes values in $\frg$ and $H$ takes values in $\frh$, according to
\begin{equation}
 \begin{aligned}
  F:=\dd_A A:=\dd A+A \wedge A\eand H:=\dd_A B:=\dd B+A \actwedge B~.
 \end{aligned}
\end{equation}
The wedge products of Lie algebra valued differential forms are defined in the obvious way: Consider $\frg$-valued forms $X_{1,2}=X^a_{1,2} \tau_a$, where $X^a_{1,2}\in \Omega^\bullet(\FR^n)$ and the $\tau_a$ are generators of $\frg$ and an $\frh$-valued form $Y=Y^a\rho_a$, where $Y^a\in \Omega^\bullet(\FR^n)$ and the $\rho_a$ are generators of $\frh$. Then
\begin{equation}
 X_1\wedge X_2:=(X^a_1\wedge X^b_2)\otimes [\tau_a,\tau_b]\eand X_1\actwedge Y:=(X^a_1\wedge Y^b)\otimes (\tau_a\acton \rho_b)~.
\end{equation}
We evidently have
\begin{equation}\label{eq:Bianchi}
 \dd_A F=0\eand \dd_A H=F\actwedge B~.
\end{equation}

It can be shown \cite{Baez:0511710}, see also \cite{Girelli:2003ev}, that a connection $(A,B)$ gives rise to well-defined parallel transport over surfaces if the so-called {\em fake curvature} vanishes:
\begin{equation}\label{eq:FakeCurvature}
 \CF:=F-\sft(B)=0~.
\end{equation}
Note that this, together with \eqref{eq:Bianchi}, implies
\begin{equation}
 \sft(H)=0\eand \dd_A H=0~.
\end{equation}

Finite gauge transformations are specified by a pair $(g,a)$ of a $\CG$-valued function $g$ and an $\frh$-valued 1-form $a$. They act according to
\begin{equation}
\begin{aligned}
 A&\rightarrow \tilde{A}:=g A g^{-1}+g \dd g^{-1}+\sft(a)~,\\
 B&\rightarrow \tilde{B}:=g\acton B +\tilde{A}\actwedge a+\dd a+ a\wedge a~.
\end{aligned}
\end{equation}
This implies
\begin{equation}
\begin{aligned}
 F&\rightarrow \tilde{F}=g F g^{-1}+\sft(\dd a)+\sft(a)\wedge \tilde{A}+\tilde{A}\wedge \sft(a)+\sft(a)\wedge \sft(a)~,\\
 H&\rightarrow \tilde{H}=g\acton H+(F-\sft(B))\actwedge a~.
\end{aligned}
\end{equation}
Due to $\sft(a)\wedge \sft(a)=\sft(a\wedge a)$, the fake curvature condition \eqref{eq:FakeCurvature} is invariant under these transformations, since
\begin{equation}
 \CF\rightarrow \tilde{\CF}=g\CF g^{-1}~.
\end{equation}
We will follow the nomenclature of e.g.\ \cite{Martins:2010ry} and refer to gauge transformations parameterized by $(g,0)$ as {\em thin} and those parameterized by $(0,a)$ as {\em fat}. In addition, we will call gauge transformations $(g,a)$ with $\sft(a)=0$ {\em ample}.

A few remarks are in order. First, as stated above, the non-abelian gerbes of Breen and Messing \cite{Breen:math0106083} are obtained when we use automorphism Lie 2-groups. Therefore, our discussion contains non-abelian gerbes, but it is more general. Second, if $\CH$ is abelian and $\acton$ and $\sft$ are trivial, we obtain the usual picture of abelian gerbes. Third, we can always use a fat gauge transformation to remove the part of $A$ that lies in the image of $\sft$. (In particular, $A$ is always pure gauge if $\sft$ is surjective.) The remaining ample gauge transformations act on $A$ as usual, and $A$ encodes the connection 1-form on a principal $\CG$-bundle. Therefore, there are covariant derivatives on the corresponding associated vector bundles\footnote{More appropriately, we would like to study associated 2-vector bundles to the principal 2-bundles we are considering. Matter fields which are sections of these 2-vector bundles should couple fully covariantly to the connection on the differential crossed
module. This, however, is beyond the scope of this letter.}. And finally, note that an M5-brane model has been recently
proposed \cite{Ho:2011ni} that uses the above language. In the following, however, we will discuss a different model built from 3-Lie algebras.

\subsection{Tensor multiplet equations of motion}

In \cite{Lambert:2010wm}, a set of equations for the fields in the non-abelian tensor multiplet in six dimensions was proposed, which are invariant under $\CN=(2,0)$ supersymmetry. The field content of the tensor multiplet, i.e.\ the self-dual 3-form field strength $h_{\mu\nu\kappa}$, the scalars $X^I$ and superpartners $\Psi$, were all assumed to take values in a 3-Lie algebra $\CA$. It was found that for the closure of the supersymmetry algebra, it was necessary to introduce an additional gauge potential taking values in the associated Lie algebra $\frg_\CA$. Moreover, a covariantly constant, $\CA$-valued vector field $C^\mu$ had to be introduced. Altogether, the proposed equations of motion read as
\begin{equation}\label{eq:eomTensor}
\begin{aligned}
 \nabla^2 X^I-\tfrac{\di}{2}[\bar{\Psi},\Gamma_\nu\Gamma^I\Psi,C^\nu]+[X^J,C^\nu,[X^J,C_\nu,X^I]]&=0~,\\
 \Gamma^\mu\nabla_\mu\Psi-[X^I,C^\nu,\Gamma_\nu\Gamma^I\Psi]&=0~,\\
 \nabla_{[\mu}h_{\nu\kappa\lambda]}+\tfrac{1}{4}\eps_{\mu\nu\kappa\lambda\sigma\tau}[X^I,\nabla^\tau X^I,C^\sigma]+\tfrac{\di}{8}\eps_{\mu\nu\kappa\lambda\sigma\tau}[\bar{\Psi},\Gamma^\tau\Psi,C^\sigma]&=0~,\\
F_{\mu\nu}-D(C^\lambda,h_{\mu\nu\lambda})&=0~,\\
\nabla_\mu C^\nu=D(C^\mu,C^\nu)&=0~,\\
D(C^\rho,\nabla_\rho X^I)=D(C^\rho,\nabla_\rho\Psi)=D(C^\rho,\nabla_\rho h_{\mu\nu\lambda})&=0~,
\end{aligned}
\end{equation}
and the supersymmetry transformations leaving these equations invariant are given by
\begin{equation}\label{eq:SUSYTensor}
 \begin{aligned}
  \delta X^I&=\di \epsb \Gamma^I\Psi~,\\
  \delta \Psi&=\Gamma^\mu\Gamma^I\nabla_\mu X^I\eps+\tfrac{1}{2\times 3!}\Gamma_{\mu\nu\lambda}h^{\mu\nu\lambda}\eps-\tfrac{1}{2}\Gamma^{IJ}\Gamma_\lambda[X^I,X^J,C^\lambda]\eps~,\\
  \delta h_{\mu\nu\lambda}&=3\di \epsb\Gamma_{[\mu\nu}\nabla_{\lambda]}\Psi+\di\epsb\Gamma^I\Gamma_{\mu\nu\lambda\kappa}[X^I,\Psi,C^\kappa]~,\\
  \delta A_\mu&=\di\epsb\Gamma_{\mu\lambda} D(C^\lambda,\Psi)~,\\
  \delta C^\mu&=0~.
 \end{aligned}
\end{equation}
Here, $(\Gamma_\mu,\Gamma_I)$, $\mu=0,\ldots,5$, $I=1,\ldots,5$, form the generators of the Clifford algebra of $\FR^{1,10}$. All the $\CA$-valued fields transform in the natural representation given by $\CA$. However, it seems impossible to consistently introduce a potential 2-form field $B$ for $h$. In \cite{Papageorgakis:2011xg}, the equations \eqref{eq:eomTensor} found a natural interpretation on loop space: The constraints on $C^\mu$ imply a factorization, $C^\mu=c^\mu C$, where $C$ is a constant element of $\CA$, and the remaining covariantly constant vector $c^\mu$ can be identified with the tangent vector to the loop. This implies that the equation $F_{\mu\nu}-D(C^\lambda,h_{\mu\nu\lambda})=0$ is very similar to a transgression, i.e.\ a map of $p+1$-forms on space-time to $p$-forms on the space of loops in space-time.

Here, however, we want to reformulate equations \eqref{eq:eomTensor} in terms of a differential crossed module $(\sft:\frh\rightarrow\frg,\acton)$. That is, we replace $\CA$ and $\frg_\CA$ by $\frh$ and $\frg$, respectively. Instead of having an extra element $C\in\CA$, we substitute all expressions $D(y,C)$, $y\in\CA$, by $\sft(y)$. Correspondingly, all 3-brackets containing $C$, i.e.\ $[y_1,C,y_2]=D(y_1,C)\acton y_2$, $y_1,y_2\in \CA$, become $\sft(y_1)\acton y_2=[y_1,y_2]$. Note that in equations \eqref{eq:eomTensor} and \eqref{eq:SUSYTensor}, $C$ appears in every 3-bracket and in every expression containing the map $D$. We will therefore obtain equations containing only the Lie structures on $\frh$ and $\frg$.

We cannot work with differential crossed modules yielding 3-Lie algebras, because in these cases, the map $\sft$ is trivial. However, we find that the equations \eqref{eq:eomTensor} e.g.\ with 3-Lie algebra $\CA=A_4$ correspond to equations using the differential crossed module $(\mt:\au(2)\rightarrow\asu(2),\acton)$ defined in section \ref{sec:NontrivialExamples}.

While the equation $F_{\mu\nu}-D(C^\lambda,h_{\mu\nu\lambda})=0$ looks like a transgression in the loop space picture, in the context of differential crossed modules it is a candidate for the fake curvature constraint \eqref{eq:FakeCurvature}. Consequently, we are led to identify $B_{\mu\nu}=h_{\mu\nu\lambda}c^\lambda$. For simplicity, we will assume $|c|>0$. Given a $B_{\mu\nu}$ satisfying $B_{\mu\nu}c^\nu=0$, we can then write
\begin{equation}\label{eq:Defh}
 h_{\mu\nu\kappa}=\frac{1}{|c|^2}\left(B_{[\mu\nu}c_{\kappa]}+\tfrac{1}{3!}\eps_{\mu\nu\kappa\lambda\rho\sigma}B^{[\lambda\rho}c^{\sigma]}\right)~,
\end{equation}
where $[\cdots]$ denotes antisymmetrization of $n$ indices with weight $1/n!$ . Note that locally and before taking gauge invariance into account, a self-dual 3-form in six dimensions has just as many components as a 2-form satisfying $B_{\mu\nu}c^\nu=0$. Such a 2-form has non-trivial components only in the five dimensional space perpendicular to $c$.

Let us now rewrite \eqref{eq:eomTensor} in the language of differential crossed modules:
\begin{equation}\label{eq:NewEomTensor}
\begin{aligned}
 \nabla^2 X^I-\tfrac{\di}{2}[\bar{\Psi},\Gamma\Gamma^I\Psi]+|c|^2[X^J,[X^J,X^I]]&=0~,\\
 \Gamma^\mu\nabla_\mu\Psi+[X^I,\Gamma\Gamma^I\Psi]&=0~,\\
 \nabla_{[\mu}h_{\nu\kappa\lambda]}+\tfrac{1}{4}\eps_{\mu\nu\kappa\lambda\sigma\tau}c^\sigma \left([X^I,\nabla^\tau X^I]+\tfrac{\di}{2}[\bar{\Psi},\Gamma^\tau\Psi]\right)&=0~,\\
H_{\mu\nu\kappa}-\tfrac{1}{3!}\eps_{\mu\nu\kappa\rho\sigma\tau}H^{\rho\sigma\tau}&=0~,\\
F_{\mu\nu}-\sft(B_{\mu\nu})&=0~,\\
\dpar_\mu c^\nu=\sft(\nabla_c X^I)=\sft(\nabla_c \Psi)=\sft(\nabla_c B_{\mu\nu})&=0~,
\end{aligned}
\end{equation}
where $\Gamma:=c^\nu\Gamma_\nu$, $\nabla_c:=c^\nu\nabla_\nu$ and $h$ is given in \eqref{eq:Defh}. Note that the commutators of spinors are to be read as commutators of the gauge structure only.

From the third equation in \eqref{eq:NewEomTensor}, we find
\begin{equation}
 c^\lambda (\nabla_{[\mu}h_{\nu\kappa\lambda]})=0~.
\end{equation}
Using this, we compute
\begin{equation}
\begin{aligned}
 H&:=\dd_A B=c^\lambda \nabla_\lambda h_{\mu\nu\kappa}\dd x^\mu\wedge \dd x^\nu\wedge \dd x^\kappa~,\\
 *H&=\tfrac{1}{3!}\eps_{\mu\nu\kappa\rho\sigma\tau}c_\lambda \nabla^\lambda h^{\rho \sigma\tau}\dd x^\mu\wedge \dd x^\nu\wedge \dd x^\kappa~,
\end{aligned}
\end{equation}
from which (together with the self-duality of $h$) we conclude that
\begin{equation}
 H=*H\eand \sft(H)=0~~~\Rightarrow~~~\sft(\nabla_c B_{\mu\nu})=0~.
\end{equation}
Thus, our definition of $B$ yields indeed a self-dual curvature 3-form. Moreover, it also answers the question why there is no potential for $h$: The field $h$ encodes the potential. And finally, note that the degrees of freedom in the gauge potential are completely determined by the 2-form potential $B$ via the fake curvature condition $F-\sft(B)=0$. Therefore, there are no additional degrees of freedom in the supermultiplet.

As we merely rewrote the equations of motion, it is clear that for certain differential crossed modules $(\sft:\frh\rightarrow\frg,\acton)$, equations \eqref{eq:NewEomTensor} are invariant under the maximal $\CN=(2,0)$ supersymmetry transformations
\begin{equation}\label{eq:newSUSY}
 \begin{aligned}
  \delta X^I&=\di \epsb \Gamma^I\Psi~,\\
  \delta \Psi&=\Gamma^\mu\Gamma^I\nabla_\mu X^I\eps+\tfrac{1}{2\times 3!}\Gamma_{\mu\nu\lambda}h^{\mu\nu\lambda}\eps-\tfrac{1}{2}\Gamma^{IJ}\Gamma [X^I,X^J]\eps~,\\
  \delta B_{\mu\nu}&=3\di \epsb\Gamma_{[\mu\nu}c^\lambda\nabla_{\lambda]}\Psi~,\\
  \delta A_\mu&=\di\epsb\Gamma_{\mu\lambda}c^\lambda \sft(\Psi)~,\\
  \delta c^\mu&=0~.
 \end{aligned}
\end{equation}
Recall that equations \eqref{eq:eomTensor} are maximally supersymmetric if the contained 3-brackets are totally antisymmetric and satisfy the fundamental identity \cite{Lambert:2010wm}. The consequences of these properties in equations \eqref{eq:eomTensor} are preserved under the rewriting $D(y,C)\rightarrow \sft(y)$, as is readily verfied. One would therefore expect that equations \eqref{eq:NewEomTensor} are invariant under the supersymmetry transformations \eqref{eq:newSUSY} for any differential crossed module $(\sft:\frh\rightarrow\frg,\acton)$. An explicit computation along the lines of \cite{Lambert:2010wm} confirms this expectation. Thus, we significantly extended the previously known examples of $\CN=(2,0)$ tensor multiplet equations.

\subsection{Comments on the tensor multiplet equations}

First of all, it is not clear to us how to make the above equations invariant under general fat gauge transformations. The equations \eqref{eq:NewEomTensor} are only invariant under thin gauge transformations $(g,0)$ with
\begin{equation}
 X^I\rightarrow \tilde{X}^I:=g\acton X^I\eand \Psi\rightarrow \tilde{\Psi}:=g\acton \Psi~.
\end{equation}
We thus recover the gauge symmetry already suggested in \cite{Lambert:2010wm}.

Second, it is nice that for $\sft$ trivial, i.e.\ the case of an abelian gerbe, $\frh$ must be abelian and the field strength $F$ necessarily vanishes. We can therefore gauge away the gauge potential and obtain a free theory:
\begin{equation}
 \dpar^2X^I=\Gamma^\mu\dpar_\mu \Psi= H-(*H)=0~.
\end{equation}

Third, we can follow \cite{Lambert:2010wm} and reduce equations \eqref{eq:NewEomTensor} to five-dimensional maximally supersymmetric Yang-Mills (mSYM) theory. For this, we dimensionally reduce along $x^5$ by imposing $\der{x^5}=0$ and fixing $c^\mu=\delta^{\mu5}g^2_{\rm YM}$. Due to $B_{\mu\nu}=h_{\mu\nu\kappa}c^\kappa$, we conclude that $B_{\mu5}=0$. This implies that $F_{\mu5}=0$ and we can therefore partially gauge fix $A_5=0$. The relation $B_{\mu 5}=0$ together with $\der{x^5}=0$ and the self-duality of $H$ also yields $H=0$. We are therefore left with the field content of mSYM theory in five dimensions. If we use the differential crossed module $(\sft:\au(N)\rightarrow\au(N),\acton)$, equations \eqref{eq:NewEomTensor} reduce to the mSYM equations with gauge algebra $\au(N)$.

As a final test, let us briefly derive the BPS equation corresponding to a (non-abelian) self-dual string. That is, we dimensionally reduce the above equations along the $x^0$- and $x^5$-directions and put $\Phi:=X^6\neq 0=X^7,\ldots,X^{10}$ as well as $H_{0ij} = H_{5ij} = 0$. Then the supersymmetry transformation of the spinors reduces to
\begin{equation}\label{eq:BPS}
 \Gamma^i\Gamma^6\nabla_i \Phi\eps+\tfrac{1}{2\times 3!}\Gamma_{ijk}h^{ijk}\eps=0~,~~~i,j,k=1,\ldots,4~.
\end{equation}
To break half of the supersymmetry, as expected for the BPS equation, we impose $\Gamma^{05}\eps=\Gamma^6\eps$ and arrive at
\begin{equation}\label{eq:sdseqn}
 h_{ijk}=\eps_{ijk\ell}\nabla^\ell \Phi\eor B_{ij}=\eps_{ijk\ell}c^k\nabla^\ell\Phi~.
\end{equation}
The fact that this equation is close but not identical to the desired $H=\dd_A B=*\dd_A \Phi$ indicates that the equations \eqref{eq:NewEomTensor} need further generalization. Note that after applying $\sft$ to both sides of equation \eqref{eq:sdseqn} and using the fake curvature constraint \eqref{eq:FakeCurvature}, we obtain
\begin{equation}
 F_{ij}=\eps_{ijk\ell}c^k\nabla^\ell\sft(\Phi)~.
\end{equation}
This should be interpreted as the Bogomolny monopole equation obtained by dimensionally reducing a self-dual string along the direction $c^k$.

Altogether, we can conclude that the 3-Lie algebra tensor multiplet equations proposed in \cite{Lambert:2010wm} can be naturally reformulated in the language of differential crossed modules while preserving $\CN=(2,0)$ supersymmetry. However, the BPS equation and issues with fat gauge transformations suggest that the thus obtained equations \eqref{eq:NewEomTensor} are not the final answer.

\subsection{M2-brane models from differential crossed modules}

Let us now come to M2-brane models. In the following, we will focus on the BLG model, but our discussion trivially extends to the ABJM model. The BLG model \cite{Bagger:2007jr,Gustavsson:2007vu} is a Chern-Simons matter theory with Lagrangian
\begin{equation}\label{eq:BLGLagrangian}
\begin{aligned}
\CL_{\rm BLG}=&\tfrac{1}{2}\lbr A, \dd A+\tfrac{1}{3} A \wedge A \rbr-\tfrac{1}{2}(\nabla_\mu X^I,\nabla^\mu X^I)+\tfrac{\di}{2}(\bar\Psi,\Gamma^\mu\nabla_\mu\Psi)\\&-\tfrac{\di}{4}(\bar \Psi,\Gamma_{IJ}[X^I,X^J,\Psi])-\tfrac{1}{6}([X^I,X^J,X^K],[X^I,X^J,X^K])~,
\end{aligned}
\end{equation}
where $(\Gamma_\mu,\Gamma_I)$, $\mu=0,1,2$, $I=1,\ldots,8$, form the Clifford algebra of $\FR^{1,10}$. The matter fields $X^I$, $I=1,\ldots, 8$ and $\Psi$ take values in a 3-Lie algebra $\CA$ with inner product $(\cdot,\cdot)$. The gauge potential $A$ takes values in the associated Lie algebra $\frg_\CA$ with invariant symmetric bilinear form $\lbr\cdot,\cdot\rbr$.

This Lagrangian is invariant under maximal $\CN=8$ supersymmetry. Replacing the 3-Lie algebra $\CA$ by a real 3-algebra preserves conformal invariance \cite{Akerblom:2009gx} and at least $\CN=2$ supersymmetry \cite{Cherkis:2008qr,Cherkis:2008ha}.

In the Lie 2-algebraic interpretation, this model has the same difficulties as the tensor multiplet equations. In particular, matter fields do not couple nicely to the gauge structure, and we will have to restrict ourselves to ample gauge transformations. But the situation is even more subtle: To extend the BLG model to a higher gauge theory, one could impose $\sft(B)=F$. However, we expect to recover the BLG model in the case $\sft=0$, for which the fake curvature condition \eqref{eq:FakeCurvature} reduces to $\CF=F-\sft(B)=F=0$. This contradicts the corresponding equation of motion of the BLG model:
\begin{equation}\label{eq:BLGFeom}
F_{\mu\nu}=\eps_{\mu\nu\lambda}(D(X^I,\nabla^\lambda X^I)+\tfrac{\di}{2}D(\bar \Psi,\Gamma^\lambda\Psi))~.
\end{equation}
Let us nevertheless explore this option a little further. First of all, the fake curvature condition requires $F$ and the right-hand side of \eqref{eq:BLGFeom} to be in the image of $\sft$, which suggests that $\sft$ should be chosen to be surjective. In this case, both gauge invariance and supersymmetry of the equations of motion are preserved, if we impose that  $\sft(B)$ and $F$ transform equally under supersymmetry transformations. We can impose the fake curvature condition by adding a Lagrange multiplier term to the action:
\begin{equation}
 \CL_1=\CL_{\rm BLG}+\lbr \Lambda,F-\sft(B)\rbr~,
\end{equation}
where $\Lambda$ is an $\frh$-valued 1-form, transforming in the adjoint of the gauge group $\frg$. Varying this new action with respect to the various fields yields the fake curvature condition, the equation \eqref{eq:BLGFeom} and the equation $\sft^*(\Lambda)=0$, which is equivalent to $\Lambda=0$ for $\sft$ surjective. Imposing that $\Lambda$ transforms trivially under supersymmetry transformations, the Lagrangian $\CL_1$ is supersymmetric on-shell.

A drawback of the Lagrangian $\CL_1$ is that for $\sft$ trivial, it does not reduce to the BLG model. One might therefore wonder, if it is sensible to introduce a $B$-field term into \eqref{eq:BLGFeom} according to
\begin{equation}\label{eq:newEOM}
\CF=F_{\mu\nu}-\mt(B_{\mu\nu})=\eps_{\mu\nu\lambda}(D(X^I,\nabla^\lambda X^I)+\tfrac{\di}{2}D(\bar \Psi,\Gamma^\lambda\Psi))~.
\end{equation}
Clearly, this choice breaks reparameterization invariance under parallel transport along surfaces (which is equivalent to the vanishing of the fake curvature). Moreover, the usual supersymmetry algebra of the BLG model does not close on-shell, unless $\sft$ is trivial \cite{Bagger:2007jr}. The equation \eqref{eq:newEOM} can be obtained from the Lagrangian
\begin{equation}
 \CL_2=\CL_{\rm BLG}-\lbr A,\sft(B)\rbr~.
\end{equation}
This yields \eqref{eq:newEOM} together with the equation $\sft^*(A)=0$, which is only gauge invariant for general $A$, if $\sft$ (and therefore $\sft^*$) is trivial. Altogether, this leads us back to differential crossed modules that are 3-algebras and thus to the BLG model.

A few final comments are in order. More general supersymmetry transformations than those induced by the ones of the BLG model might allow for unknown examples of maximally supersymmetric Chern-Simons matter theories. A dimensional analysis suggests that one cannot trivially include terms involving the $B$-field into the supersymmetry transformations of the BLG model. However, given an additional covariantly constant vector $c$ similar to that appearing in the M5-brane equations, this may be possible.

As far as a unification of M2- and M5-brane models is concerned, it might be interesting to note that the right-hand side of \eqref{eq:BLGFeom} also appears in the equation for $h_{\mu\nu\kappa}$ contained in \eqref{eq:NewEomTensor}.

The BPS equation, i.e.\ the generalized Basu-Harvey equation, is obtained by dimensionally reducing the condition that the supersymmetry transformations of $\Psi$ vanish. As the $B$-field does not appear, it is essentially identical to the original Basu-Harvey equation:
\begin{equation}
 \dder{s} X^\mu=\tfrac{1}{3!}\eps^{\mu\nu\kappa\lambda}D(X^\nu,X^\kappa)\acton X^\lambda~.
\end{equation}

A clearer interpretation of the BLG model in the context of higher gauge theories is desirable. We suspect that this issue, together with the problem of coupling higher gauge theories to matter fields, will only be resolved by considering matter fields as sections of 2-vector bundles.

\section*{Acknowledgments}

We would like to thank David Berman, Neil Lambert and Yutaka Matsuo for discussions. We are particularly grateful to Martin Wolf for discussions and comments on a first draft of this paper. We would also like to thank the organizers of the program ``Mathematics and Applications of Branes in String and M-theory'' at the Newton Institute, Cambridge, during which the major part of this work was completed. This work was supported by a Career Acceleration Fellowship from the UK Engineering and Physical Sciences Research Council.



\pdfbookmark[0]{References}{r1}

\end{document}